\documentclass[10pt, conference]{IEEEtran}
\IEEEoverridecommandlockouts
\newcommand{\bfh}{{\boldsymbol h}}

\newcommand{\diag}{\mathrm{diag}}

\newcommand{\bfr}{{\boldsymbol r}}

\newcommand{\bfv}{{\boldsymbol v}}
\newcommand{\bfw}{{\boldsymbol w}}

\newcommand{\bfG}{{\boldsymbol G}}

\newcommand{\bfT}{{\boldsymbol T}}
\newcommand{\bfU}{{\boldsymbol U}}

\newcommand{\bfW}{{\boldsymbol W}}

\newcommand{\Phimat}{{\boldsymbol{\it{\Phi}}}} %\newcommand{\Phimat}{\mathbf{\Phi}}
\newcommand{\phivec}{\pmb{\varphi}} %\newcommand{\phivec}{\pmb{\phi}}

\newcommand{\dm}{\mathrm{d}}
\newcommand{\Tm}{\mathrm{T}}

\newcommand{\lam}{{\boldsymbol{\lambda}}}

\newcommand{\Hm}{{\mathrm{H}}}

\newcommand{\phivecr}{ \phivec_{\text{r}}} 
\newcommand{\phivect}{ \phivec_{\text{t}}}

\newcommand{\rhor}{ \boldsymbol{\rho}_{\text{r}}}
\newcommand{\rhot}{ \boldsymbol{\rho}_{\text{t}}}

\newcommand{\phir}{ \phi_{\text{r},n}} 
\newcommand{\phit}{ \phi_{\text{t},n}}
\newcommand{\rt}{ \text{r}}
\newcommand{\ttxt}{ \text{t}}

\newcommand{\thetar}{ {\boldsymbol{\theta}}_{\text{r}}} 
\newcommand{\thetat}{ {\boldsymbol{\theta}}_{\text{t}}}
\newcommand{\Ocal}{\mathcal{O}}
\def\BibTeX{{\rm B\kern-.05em{\sc i\kern-.025em b}\kern-.08em T\kern-.1667em\lower.7ex\hbox{E}\kern-.125emX}}
\usepackage{cite}
\usepackage{amsmath,amssymb,amsfonts}
\usepackage{graphicx}
\usepackage{textcomp}
\usepackage{xcolor}
\def\BibTeX{{\rm B\kern-.05em{\sc i\kern-.025em b}\kern-.08em
    T\kern-.1667em\lower.7ex\hbox{E}\kern-.125emX}}

\usepackage{afterpage}
\usepackage{tikz}
\usepackage{lipsum}
\usepackage{fancyhdr}

\usetikzlibrary{%
	arrows,%
	automata,%
	backgrounds,%
	calc,%
	patterns,%
	plotmarks,%
	positioning,%
	shapes.geometric,%
	shapes,%
	snakes,%
	decorations.pathmorphing,%
	decorations.shapes,%
	graphs,%
	chains,%
	arrows.meta, %
}

\usepackage{cite}
\usepackage{amsmath,amssymb,amsfonts}
\usepackage{lipsum}
\usepackage{graphicx}
\usepackage{subcaption}
\usepackage{hyperref}
\usepackage{textcomp}
\usepackage{xcolor}
\def\BibTeX{{\rm B\kern-.05em{\sc i\kern-.025em b}\kern-.08em T\kern-.1667em\lower.7ex\hbox{E}\kern-.125emX}}

\usepackage{tikz}
\usepackage{pgfplots}
\pgfplotsset{compat=newest}
\pgfplotsset{plot coordinates/math parser=false}
\newlength\figH  %figureheight
\newlength\figW  %figurewidth
\usetikzlibrary{quotes,arrows.meta}
\usepackage{physics}
\usepackage{amssymb}
\usepackage{amsmath}
\usepackage{amsthm}
\usepackage{float}
\usepackage{algorithm}
\usepackage{algpseudocode}
\usepackage{algorithm,caption}
\usepackage[english]{babel}

\usepackage{graphicx}
\usepackage{wrapfig}
\usepackage{caption}
\usepackage{array}
\usepackage{multirow}
\usepackage{arydshln}
\usepackage{color}
\usepackage{xcolor}
\usepackage{tcolorbox}
\usepackage[framemethod=tikz]{mdframed}
\fancypagestyle{IEEEtitlepagestyle}{
    \fancyhf{} % clear all header and footer fields
    \fancyfoot[L]{\vspace{12pt}%
    \footnotesize
    \copyright This work has been submitted to the IEEE for possible publication. Copyright may be transferred without notice, after which this version may no longer be accessible.}
}
\begin{document}
\title{Sum-Rate Optimisation of a Multi-User STAR-RIS-Aided System with Low Complexity\\
%{\footnotesize \textsuperscript{*}Note: Sub-titles are not captured for https://ieeexplore.ieee.org  and
%should not be used}
%\thanks{This work was supported by the Federal Ministry of Education and Research of Germany in the programme of “Souverän. Digital. Vernetzt.”. Joint project 6G-life, project identification number: 16KISK002.}
}

\author{\IEEEauthorblockN{Sadaf Syed, Wolfgang Utschick, Michael Joham}
\IEEEauthorblockA{{School of Computation, Information and Technology, Technical University of Munich, Germany}\\
Emails: \{sadaf.syed, utschick, joham\}@tum.de}
}

%\author{\IEEEauthorblockN{Sadaf Syed}
%\IEEEauthorblockA{\textit{dept. name of organization (of Aff.)} \\
%\textit{name of organization (of Aff.)}\\
%City, Country \\
%email address or ORCID}
%\and
%\IEEEauthorblockN{Michael Joham}
%\IEEEauthorblockA{\textit{dept. name of organization (of Aff.)} \\
%\textit{name of organization (of Aff.)}\\
%City, Country \\
%email address or ORCID}
%\and
%\IEEEauthorblockN{Wolfgang Utschick}
%\IEEEauthorblockA{\textit{dept. name of organization (of Aff.)} \\
%\textit{name of organization (of Aff.)}\\
%City, Country \\
%email address or ORCID}

%}

\maketitle

\begin{abstract}
Reconfigurable intelligent surface~(RIS) is a promising technology for future wireless communication systems. However, the conventional RIS can only reflect the incident signal. Hence, it provides a limited coverage, as compared to a simultaneously transmitting and reflecting RIS (STAR-RIS). Prior works on the STAR-RIS address the power minimisation or the sum-rate maximisation problem by reformulating the objective problem as a convex optimisation problem and then employing numerical tools like CVX to obtain the solution, which introduces significant computational complexity leading to a huge runtime, making the algorithms impractical for real-world implementation. In this paper, we propose a low complexity solution for the optimisation of a multi-user STAR-RIS system, where the non-convex optimisation problem is decomposed into multiple convex sub-problems with closed-form optimal solutions. The simulation results illustrate that our proposed algorithm achieves similar performance to CVX-based solutions in the literature while being computationally efficient. 
\end{abstract}


\section{Introduction}
Reconfigurable intelligent surface~(RIS), which is a prospective low-cost technology, has received considerable attention for B5G/6G systems \cite{wu2021intelligent,zhang2020capacity, wu2019intelligent, guo2020weighted, semmler2022linear, syed2023design, syed2024design}. Specifically, an RIS is composed of a large number of reflecting elements, where the phase shifts can be configured to achieve the desired system's performance. RIS consists of a large number of lossless reconfigurable reflecting elements that can reflect the incident signal in a controlled manner to achieve the desired performance objectives. Additionally, the energy and hardware costs of the lossless elements of the RIS are significantly lower compared to the traditional active antennas at the base station (BS). Thus, they can be scaled much more easily than the antennas at the BS. Most existing algorithms consider reflecting-only RIS, which can only reflect the incident signals \cite{wu2021intelligent,zhang2020capacity, wu2019intelligent, guo2020weighted, semmler2022linear, syed2023design, syed2024design}. This means that the conventional RIS can only serve the users located on the same side as the BS. This geographical constraint limits the flexibility and performance gains, as the users are typically located on both sides of the RIS. To overcome this limitation, the concept of simultaneously transmitting and reflecting RIS (STAR-RIS) was proposed in \cite{mu2021simultaneously, xu2021star, niu2021weighted}, where the incident wireless signal on each element of the STAR-RIS is split into reflecting and transmitting components. The transmitted and reflected signals can be reconfigured via independent coefficients by manipulating the electric and magnetic currents of each element \cite{xu2021star}. 
Compared to the conventional RISs, STAR-RIS requires optimisation of the amplitude coefficients of each element in addition to the phase shifts. The amplitude coefficients determine the percentage of energy allocated to reflection and transmission modes, which are coupled due to energy conservation laws. This makes the optimisation problem of STAR-RIS very challenging. The existing works on STAR-RIS like \cite{mu2021simultaneously, niu2021weighted} employ interior-point solvers like CVX \cite{grant2014cvx} to obtain the optimal passive beamforming vectors of the STAR-RIS. However, this results in high computational complexity and runtime, making the algorithms infeasible for real-world implementation. In this work, we firstly reformulate the problem to a tractable form using the fractional programming (FP) approach \cite{fractional1, fractional2}. We propose a low complexity iterative solution where the phase shifts and the amplitude coefficients of the STAR-RIS are solved in closed-form, which leads to much lower computational complexity and smaller runtime as compared to the CVX-based algorithms of \cite{mu2021simultaneously, niu2021weighted}.
\section{System Model}
We consider the downlink (DL) of a multi-user STAR-RIS-aided system where the BS is equipped with $M$ antennas and the STAR-RIS consists of $N$ lossless elements. The wireless signal from the BS that is incident on a single element of the STAR-RIS is divided into a transmitting and a reflecting signal. We consider $K$ single-antenna users where $K_{\rt}$ users lie on the reflecting side and $K_{\ttxt}$ users lie on the transmitting side of the STAR-RIS, such that $K_{\rt} + K_{\ttxt} = K$. For each element of the STAR-RIS, the phase shifts for reflection and transmission can be designed independently \cite{xu2021star}. However, the amplitude coefficients of the reflection and transmission parts are coupled by the law of energy conservation since the elements of the STAR-RIS are lossless. The reflection and transmission coefficient matrices of the STAR-RIS are given by
\begin{align}
   \Phimat_{\rt} &=  \text{diag}\left(\sqrt{\rho_{\rt,1}}\theta_{\rt, 1},\cdots, \sqrt{\rho_{\rt,N}}\theta_{\rt, N} \right) \\
   \Phimat_{\ttxt} & =\text{diag}\left( \sqrt{\rho_{\ttxt,1}}\theta_{\ttxt, 1},\cdots, \sqrt{\rho_{\ttxt,N}}\theta_{\ttxt, N}\right) 
\end{align}
where the phase shifts for the reflection and transmission for the $n$-th element are denoted by $\theta_{\rt, n}$ and $ \theta_{\ttxt, n}$ respectively, with $\sqrt{\rho_{\rt,n}}$ and $\sqrt{\rho_{\ttxt,n}}$ being the corresponding amplitude coefficients with $\rho_{\rt,n}, \:\: \rho_{\ttxt,n} \in [0,1] $. Additionally, we have $|\theta_{\rt / \ttxt, n}| = 1 \:\forall \:n$, and $\phivec_{\rt / \ttxt} = \diag ( \Phimat_{\rt / \ttxt})$ denotes the corresponding vector. Moreover, we have the constraint that the energy of the incident signal must be equal to the sum of the energies of the reflecting and transmitting signals, i.e.,
\begin{align}
    \rho_{\rt,n} + \rho_{\ttxt,n} = 1 \:\forall \:n . \label{eqn3}
\end{align} 
As it can be seen from \eqref{eqn3}, one can control the amount of signal reflected and transmitted by each element of the STAR-RIS by adjusting the amplitude coefficients $\rho_{\rt,n} \:\: \text{and} \:\: \rho_{\ttxt,n}$. Three operating protocols for STAR-RIS have been proposed in the literature \cite{mu2021simultaneously, niu2021weighted}.
\begin{enumerate}
    \item Energy Splitting (ES): In this protocol, all elements of the STAR-RIS simultaneously operate in reflecting and transmitting modes with $\rho_{\rt,n}, \:\: \rho_{\ttxt,n} \in [0,1] $ and $\rho_{\rt,n} + \rho_{\ttxt,n} = 1 \:\forall \:n$. 
    \item Mode Splitting (MS): In this configuration, some elements of the STAR-RIS fully operate in the reflecting mode and the others fully operate in the transmitting mode, i.e., $\rho_{\rt,n}, \:\: \rho_{\ttxt,n} \in \{0,1\}$. If $N_{\rt}$ denotes the number of elements operating in the reflecting mode with $\rho_{\rt} = 1$, $\rho_{\ttxt} = 0$, and $N_{\ttxt}$ denotes the number of elements operating in the transmitting mode with $\rho_{\rt} = 0$, $\rho_{\ttxt} = 1$, we have $N_{\rt} + N_{\ttxt} = N$. 
    \item Time Splitting (TS): In this protocol, all elements of the STAR-RIS alternatingly operate in full transmitting or reflecting modes in different time slots. If $\delta \in [0,1]$ denotes the percentage of time allocated for the reflecting mode and $1 - \delta $ is the fraction of time for the transmitted mode, the algorithm needs to optimise $\delta$ to achieve the optimal performance.    
\end{enumerate}
The coupling of the reflecting and transmitting parts due to the amplitude coefficients makes the optimisation of ES and MS modes of STAR-RIS complicated. The TS mode is easier to optimise because of the decoupled reflecting and transmitting modes in each time slot. However, it requires strict time synchronisation, which makes its hardware implementation more complicated. Additionally, since the TS mode does not serve all the users in the same time slot, its gain is much lower compared to the ES and MS modes where all the users are simultaneously served. It is to be noted here that the MS protocol is a subset of the ES protocol containing only the binary values of the amplitude coefficients. Hence, the MS protocol cannot achieve the similar gain as the ES protocol, and so, we restrict our discussion to ES and TS modes in this paper.  
The direct channel from the BS to the $k$-th user is denoted by $\bfh_{\dm, k}\in \mathbb{C}^{M\times 1}$, that from the RIS to the $k$-th user is denoted by $\bfr_k\in \mathbb{C}^{N \times 1}$ and the channel from the BS to the RIS is denoted by $\bfT$. The effective channel for the $k$-th user is given by
\begin{align}
\label{eqn4.1}
\bfh_{k}^{\Hm} & = \bfh_{\dm,k}^{\Hm} + \bfr_{k}^{\Hm} \Phimat_k^{\Hm}\bfT =  \bfh_{\dm,k}^{\Hm} +  \phivec_k^{\Hm}\bfG_{k}
\end{align}
where $\bfG_k = \diag(\bfr_k^{\Hm})\bfT  \in \mathbb{C}^{N \times M}$, $\phivec_k = \diag(\Phimat_k) \in \{\phivec_{\rt}, \phivec_{\ttxt} \}$ depending on the location of the $k$-th user if it is on the reflecting or on the transmitting side of the STAR-RIS.
The total received signal at the $k$-th user is given by
\begin{align}
    y_{k} = \bfh_{k}^{\Hm}\sum\limits_{i = 1 }^K \bfw_{i}\:s_{i} + z_{k}
\end{align}
where $z_{k} \sim\mathcal{N_\mathbb{C}}(0, \sigma_k^2) $ denotes the additive white Gaussian noise (AWGN) at the $k$-th user's side and $\bfw_k \in \mathbb{C}^{M} $ is the corresponding beamforming vector.
The rate of the $k$-th user is given by $\mathrm{log}_2(1 + \gamma_{k})$, where $\gamma_{k}$ denotes the SINR of the $k$-th user, which is given by
\begin{align}
\label{4.3}
 \gamma_{k} = \frac{\big|{\bfh_{k}^{\Hm}}{\bfw_{k}}\big|^2}{\sum\nolimits_{\substack{i = 1 ,i \neq k}}^{K}\big|{\bfh_{k}^{\Hm}}{\bfw_{i}}\big|^2 + \sigma_k^2}. 
\end{align}
The objective is to maximise the sum-rate of the users with respect to the active beamforming at the BS and the passive beamforming by the STAR-RIS. 
\section{Optimisation Problem and Solution Approach}
In this section, we develop the optimisation problems for both ES and TS modes, and derive their low-complexity solutions. 
\subsection{Energy Splitting (ES) Mode}
The sum-rate maximisation problem for the ES mode is 
\begin{subequations}
\begin{alignat}{2}
&\!\max\limits_{\bfW,\thetar, \thetat, \rhor, \rhot}      &\quad& \sum\nolimits_{k=1}^{K}\log_2(1 + \gamma_k)  \tag{P1} \label{eqn4.4}\\
&\text{subject to} &      &  \sum\nolimits_{k=1}^{K}\norm{\bfw_k}^2 \leq P_\text{t}, \label{eqn4.5}\\
&                  &      & |\theta_{\rt,n}| = 1,  \quad |\theta_{\ttxt,n}| = 1\:\forall \:n \label{eqn4.6} \\
&                  &      & \rho_{\rt,n} + \rho_{\ttxt,n} = 1, \quad 0 \leq \rho_{\rt,n}, \rho_{\ttxt,n} \leq 1 \:\forall \:n \label{eqn4.7}
\end{alignat}
\end{subequations}
where $\bfW = [\bfw_1,\cdots, \bfw_K]^{\Tm}$, $\rhor = [\rho_{\rt,1},\cdots, \rho_{\rt,N}]^{\Tm}$, $\rhot = [\rho_{\ttxt,1},\cdots, \rho_{\ttxt,N}]^{\Tm}$, $\thetar = [\theta_{\rt,1},\cdots, \theta_{\rt,N}]^{\Tm}$, $\thetat = [\theta_{\ttxt,1},\cdots, \theta_{\ttxt,N}]^{\Tm}$. 
The objective function is firstly reformulated into a tractable form using the FP approach in 2 steps with the help of auxiliary variables, as done in \cite{fractional1, fractional2}. \\
1) {\it{Lagrangian Dual Transform}}: By introducing auxiliary variables $\lam = [\lambda_1, \cdots, \lambda_K]^{\Tm}$, the sum of logarithm function can be taken care of as follows:
\begin{align*}
%\label{4.6}
    \log(1 + \gamma_k) = \: \!\max\limits_{\lambda_k \geq 0} \: \log(1 + \lambda_k) - \lambda_k + \dfrac{(1 + \lambda_k)\gamma_k}{1 + \gamma_k}
\end{align*}
where the expression in the LHS is equivalent to that in the RHS for the optimal solution $\lambda_k = \gamma_k$. \\ 
2) {\it{Quadratic Transform}}: Given $\lambda_k$, in the second step, the fractional term is simplified using the auxiliary variables ${\boldsymbol{\beta}} = [\beta_1, \cdots, \beta_K]^{\Tm}$.
\begin{align}
%\label{qt}
\sum\limits_{k=1}^{K} \dfrac{(1 + \lambda_k)\gamma_k(\bfw_k)}{1 + \gamma_k(\bfw_k)}& = \!\max\limits_{ {{\beta_k}}} \sum\limits_{k=1}^{K}2  \sqrt{1 + \lambda_k}\text{Re}\big\{\beta_k^{*}\bfh_k^{\Hm}\bfw_k\big\} \nonumber \\ & - \sum\nolimits_{k=1}^{K}|\beta_k|^2 \left({\sum\limits_{{i = 1}}^{K}\big|{\bfh_{k}^{\Hm}}{\bfw_{i}}\big|^2 + \sigma_k^2} \right). \nonumber
\end{align}
With these simplification steps of the FP approach, one can adopt the iterative non-convex block coordinate descent~(BCD) method \cite{bcd, bcdmain} to alternatingly optimise the auxiliary variables, active beamforming vectors at the BS, and the passive beamforming vectors at the STAR-RIS. The closed-form solution of the auxiliary variables alternatingly maximising the objective function in \eqref{eqn4.4} reads as \cite[Sec. IV]{syed2024design}
\begin{align}
    \lambda_k = {\gamma}_{k}, \quad   \beta_k =  \dfrac{\left(\sqrt{1 + \lambda_k}\right)\bfh_k^{\Hm} \bfw_k}{{\sum\limits_{{i = 1}}^{K}\big|{\bfh_{k}^{\Hm}}{\bfw_{i}}\big|^2 + \sigma_k^2}} .  \label{betak1}
\end{align}
Optimisation of $\bfW$ also requires satisfying the DL power constraint at the BS, and it can be efficiently handled by applying the rescaling trick mentioned in \cite{joham2002transmit, christensen2008weighted, zhao2023rethinking}. The idea is to introduce the constraint into the objective function itself by exploiting the fact that the inequality constraint in \eqref{eqn4.5} will be satisfied with equality for the optimal precoding filters (see \cite[Sec. IV]{syed2024design}). With $\sum\nolimits_{k=1}^{K}\norm{\bfw_k}^2 = P_t$, thus, $\sigma_k^2 = \sigma_k^2 \sum\nolimits_{i=1}^{K}\norm{\bfw_i}^2/P_t$, the optimisation problem for the update of the precoders is given by
\begin{align}
    \!\max\limits_{\bfW} & \sum\limits_{k=1}^{K}2  \sqrt{1 + \lambda_k}\text{Re}\big\{\beta_k^{*}\bfh_k^{\Hm}\bfw_k\big\} \nonumber \\ &- \sum\limits_{k=1}^{K}|\beta_k|^2 \left({\sum\limits_{{i = 1}}^{K}\big|{\bfh_{k}^{\Hm}}{\bfw_{i}}\big|^2 + \sum\limits_{{i = 1}}^{K}\dfrac{\sigma_k^2}{P_t}\bfw_i^{\Hm}\bfw_i} \right) \label{obj}
\end{align}
and its optimal solution can be expressed as
\begin{align}
    \bfw_k = {\beta}_k \sqrt{1 + {\lambda}_k}\left(\sum\limits_{i = 1}^{K}\dfrac{\big|{\beta}_{i}\big|^2}{P_t} \sigma_i^2\mathbf{I}_M + \sum\limits_{i = 1}^{K}\big|{\beta}_{i}\big|^2{\bfh}_i{\bfh}_{i}^{\Hm}\right)^{-1}{\bfh}_k. \label{solw}
\end{align}
Even though the DL power constraint has been incorporated into the objective function of \eqref{obj}, the optimal solution in \eqref{solw} must be rescaled in the end to ensure that the constraint is satisfied with equality (cf.~\cite{joham2002transmit, christensen2008weighted, zhao2023rethinking}). \\
Now, considering the optimisation of the passive beamforming vectors of the STAR-RIS, we firstly express the objective as a function of $\thetar$, $\thetat$, $\rhor$ and $\rhot$, keeping the other variables fixed and dropping the irrelevant constant terms.  
\begin{subequations}
\begin{alignat}{2}
   &\!\max\limits_{\phivecr, \phivect} \:      2 \text{Re}\big \{\phivecr^{\Hm} \bfv_{\rt} \big \} - \phivecr^{\Hm} \bfU_{\rt}\phivecr + 2 \text{Re}\big \{\phivect^{\Hm} \bfv_{\ttxt} \big \} - \phivect^{\Hm} \bfU_{\ttxt}\phivect \label{P2} \tag{P2} \\       &\text{s.t.} \quad \eqref{eqn4.6}, \eqref{eqn4.7}     \nonumber   %              |\theta_{\rt,n}| = 1,  \quad |\theta_{\ttxt,n}| = 1, \quad \rho_{\rt,n} + \rho_{\ttxt,n} = 1, 0 \leq \rho_{\rt,n}, \rho_{\ttxt,n} \leq 1 \:\forall \:n \label{P2} \tag{P2} 
   \end{alignat}
\end{subequations}
where $\bfU_{\rt/ \ttxt}$ and $\bfv_{\rt/ \ttxt}$ are given by
\begin{align}
    \bfU_{\rt/ \ttxt} & = \sum\nolimits_{k = 1}^{K_{\rt} / K_{\ttxt}} \big|{\beta_k^{\rt/ \ttxt}} \big|^2 \sum\nolimits_{i = 1}^{K} \bfG_k^{\rt/ \ttxt}{\bfw}_i {\bfw}_i^{\Hm} (\bfG_k^{\rt/ \ttxt})^{\Hm}
    \end{align}
    \begin{align}
    \bfv_{\rt/ \ttxt} & = \sum\nolimits_{k = 1}^{K_{\rt} / K_{\ttxt}} {(\beta_k^{{ \rt/ \ttxt}})^{*}} \sqrt{1 + {\lambda_k^{\rt/ \ttxt}}}  \bfG^{\rt/ \ttxt}_k{\bfw}^{\rt/ \ttxt}_k   
    \nonumber \\ &- \sum\nolimits_{k = 1}^{K_{\rt} / K_{\ttxt}} \big|{\beta}^{\rt/ \ttxt}_k \big|^2  \sum\nolimits_{i = 1}^{K}\bfG_k^{\rt/ \ttxt}{\bfw}_i{\bfw}_i^{\Hm}\bfh^{\rt/ \ttxt}_{\dm,k} 
    \end{align}
    respectively. The unit modulus constraints in \eqref{P2} make the optimisation problem non-convex and hence, its optimal solution in closed-form is difficult to compute. The above sub-problem for the passive beamforming of the STAR-RIS is solved in the literature \cite{mu2021simultaneously, niu2021weighted} using the CVX solver \cite{grant2014cvx}, which makes the algorithms computationally very expensive because of the high runtime of the CVX (as demonstrated later in the paper). To address this issue, we perform the element-wise optimisation of the passive beamforming vectors where each of the $N$ elements is successively optimised keeping the others fixed. The objective function can be once again reformulated as a function of $\theta_{\rt,n}$, $\theta_{\ttxt,n}$, $\rho_{\rt,n}$ and $\rho_{\ttxt,n}$ keeping the other $N-1$ elements fixed, yielding
\begin{subequations}
\begin{alignat}{2}
   &\!\max\limits_{\phir, \phit} \: 2 \sqrt{\rho_{\rt,n}}\text{Re}\big \{\theta_{\rt,n}^{*} a_{\rt, n} \big \} + 2 \sqrt{\rho_{\ttxt,n}}\text{Re}\big \{\theta_{\ttxt,n}^{*} a_{\ttxt, n} \big \} \nonumber \\ & \quad \quad - \rho_{\rt,n}u_{\rt,n,n}  - \rho_{\ttxt,n}u_{\ttxt,n,n} \tag{P3} \label{P3}\\       &\text{s.t.}\:                    |\theta_{\rt,n}| = 1,  |\theta_{\ttxt,n}| = 1,  \rho_{\rt,n} + \rho_{\ttxt,n} = 1,  0 \leq \rho_{\rt,n}, \rho_{\ttxt,n} \leq 1 \:\forall \:n  \nonumber
   \end{alignat}
\end{subequations} 
\begin{align}
   \text{where} \quad a_{\rt, n} = v_{\rt,n} - \sum\nolimits_{\substack{i = 1 ,i \neq n}}^{N} \sqrt{\rho_{\rt,i}}\theta_{\rt,i} u_{\rt,n,i} \label{15}
\end{align}
\begin{align}
    a_{\ttxt, n} = v_{\ttxt,n} - \sum\nolimits_{\substack{i = 1, i \neq n}}^{N} \sqrt{\rho_{\ttxt,i}}\theta_{\ttxt,i} u_{\ttxt,n,i} \label{16}
\end{align}
with $u_{\rt/\ttxt,i,j} = \bfU_{\rt / \ttxt}(i,j)$. Note that $a_{\rt, n}$ and $a_{\ttxt, n}$ are constants w.r.t. the coefficients of the $n$-th element of the STAR-RIS $\rho_{\rt/\ttxt,n}$ and $\theta_{\rt/\ttxt,n}$. It is evident from \eqref{P3} that the optimisation of $\theta_{\rt,n}$ and $\theta_{\ttxt,n}$ are decoupled from each other, and their closed-form optimal solutions are given by
\begin{align}
    \theta_{\rt,n} = e^{j \angle a_{\rt, n}} \quad
    \theta_{\ttxt,n} = e^{j \angle a_{\ttxt, n}}. \label{eqn27}
\end{align}
After optimising $ \theta_{\rt,n}$ and $\theta_{\ttxt,n}$, we can maximise \eqref{P3} w.r.t. $\rho_{\rt,n}$ and $\rho_{\ttxt,n}$. Denoting $\rho_{\rt,n}$ as $x$, the optimisation problem for the update of transmitting and reflecting coefficients reads as
\begin{align}
    &\!\max\limits_{x}      & &  \sqrt{x}b_{\rt,n} + \sqrt{1-x}\:b_{\ttxt,n} - x \:U_n  \quad \text{s.t.}\quad 0 \leq x \leq 1 \label{P4} \tag{P4}
\end{align}
where $b_{\rt, n} = 2 \text{Re}\{\theta_{\rt,n}^{*} a_{\rt, n} \}$, $b_{\ttxt, n} = 2 \text{Re}\{\theta_{\ttxt,n}^{*} a_{\ttxt, n} \}$ and $U_n =  u_{\rt,n,n} -u_{\ttxt,n,n}$ . \\
{\it{\bf{Theorem 1}}}: For the optimal phase shifts of the $n$-th element of STAR-RIS given by \eqref{eqn27}, the function
\begin{align}
    f(x) = \sqrt{x}b_{\rt,n} + \sqrt{1-x}\:b_{\ttxt,n} - x\:U_n \label{theorem1}
\end{align}
is concave in $x$, and hence, the global optimal solution of $\eqref{P4}$ can be computed for the given optimal $\theta_{\rt,n}$ and $\theta_{\ttxt,n}$.
\begin{proof}
With the expression of $f(x)$ in \eqref{theorem1}, its second derivative w.r.t. $x$ is given by
\begin{align}
    f''(x) = -\dfrac{1}{4}x^{-3/2}b_{\rt,n}  -\dfrac{1}{4}(1-x)^{-3/2}b_{\ttxt,n} \label{29}
\end{align}
For the optimal values of $\theta_{\rt,n}$ and $\theta_{\ttxt,n}$ given by \eqref{eqn27}
\begin{align}
    b_{\rt, n} = 2 \: |\theta_{\rt,n}| |a_{\rt, n}| =  2 \:|a_{\rt, n}| \geq 0 \\
    b_{\ttxt, n} = 2 \: |\theta_{\ttxt,n}| |a_{\ttxt, n}| = 2 \: |a_{\ttxt, n}| \geq 0. 
\end{align}
Since $ 0 \leq x \leq 1$, $f''(x) \leq 0$ and this makes $f(x)$ a concave function of $x$.  
\end{proof}
In order to maximise $f(x)$ w.r.t. $x$, we compute the solution of $f'(x) = 0$, which gives $g(x) = 0$, where
{%\small
\begin{align}
    g(x) =  \dfrac{1}{2}x^{-1/2}b_{\rt,n}  -\dfrac{1}{2}(1-x)^{-1/2}b_{\ttxt,n}-U_n. %= 0.
\end{align}
}
{\it{\bf{Corollary}}}: $g(x)$ is monotonically decreasing in $x$, and hence, the solution of $g(x) = 0$ can be computed by 1-dimensional bisection search over the range of $x \in [0,1]$.
\begin{proof}
$g'(x) = f''(x)$ [cf.\eqref{29}], and from Theorem~1, $f''(x) \leq 0$ for the optimal values of $\theta_{\rt,n}$ and $\theta_{\ttxt,n}$.
\end{proof}
Hence, \eqref{eqn4.4} can be solved using the BCD method by iteratively optimising the auxiliary variables, precoders, phase shifts, and amplitude coefficients of each element of the STAR-RIS until convergence. The convergence of the BCD method is established in \cite{bcd, bcdmain}, and since we achieve the optimal solution of every sub-problem, the proposed algorithm will converge to a stationary point. 
\subsection{Time Splitting (TS) Mode}
TS mode serves reflecting and transmitting users in different time slots, so its optimisation is decoupled and similar to the reflecting-only RIS. However, this approach needs optimisation of the fraction $\delta$ in the end, which determines the amount of allotted time for each mode. The optimisation of auxiliary variables, $\bfW$, and the phase shifts are similar to the ES mode. The optimisation problem for the update of $\delta$ is given by
\begin{subequations}
\begin{alignat}{2}
&\!\max\limits_{\delta}      &\quad& \delta\sum\nolimits_{k=1}^{K_{\rt}}\log_2(1 + \gamma^{\rt}_k)  + (1 - \delta )\sum\nolimits_{k=1}^{K_{\ttxt}}\log_2(1 + \gamma^{\ttxt}_k) \nonumber\\
&     \text{s.t.}          &      &  0 \leq \delta \leq 1 \tag{P5} \label{P5} \end{alignat}
\end{subequations}
where $\sigma_k^2$ in $\gamma_k$ is scaled by $\delta$ to ensure a fair comparison with the ES protocol (cf. \cite{mu2021simultaneously}). The objective function can be reformulated using the FP approach as $t(\delta) = a \:\delta - b\: \delta^2$, where $a, b$ are provided in Appendix. \\
{\it{\bf{Proposition 1}}}: $t(\delta)$ is concave in $\delta$ and it is monotonically decreasing in $\delta$, and hence the solution of $t'(\delta) = 0$ can be computed by the bisection method over the range of $\delta \in [0,1]$.
\begin{proof}
See Appendix.
\end{proof}
\subsection{Complexity of the Algorithm}
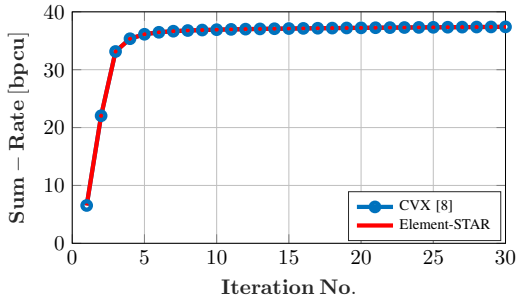
\begin{figure}		
		\scalebox{0.80}{\definecolor{mycolor1}{rgb}{0.00000,0.44700,0.74100}%
\definecolor{mycolor2}{rgb}{0.85000,0.32500,0.09800}%
\begin{tikzpicture}

\begin{axis}[%
width=0.9\figW,
height=\figH,
at={(0\figW,0\figH)},
scale only axis,
xmin=0,
xmax=30,
xlabel style={font=\color{white!15!black}},
xlabel={$\bf{Iteration\:No.}$},
ymin=0,
ymax=40,
ylabel style={font=\color{white!15!black}},
ylabel={$\bf{Sum-Rate\:[bpcu]}$},
axis background/.style={fill=white},
xmajorgrids,
ymajorgrids,
legend style={at={(0.99,0.01)}, anchor=south east, legend cell align=left, align=left, draw=white!15!black, row sep=-0.05cm, font = \scriptsize}
]
\addplot [color=mycolor1, solid, line width=2.0pt, mark=o, mark options={solid, mycolor1}]
  table[row sep=crcr]{%
1	6.53835880665336\\
2	22.0418635892511\\
3	33.1477406147662\\
4	35.3521846944333\\
5	36.1211103193523\\
6	36.4642358294226\\
7	36.6505904886127\\
8	36.7679774754395\\
9	36.85052498257\\
10	36.9134580862667\\
11	36.9643458974088\\
12	37.0072943473312\\
13	37.0446953431044\\
14	37.0780246619482\\
15	37.1082470408301\\
16	37.1360091408594\\
17	37.161769892826\\
18	37.1858636136321\\
19	37.2085410520505\\
20	37.2299951909837\\
21	37.2503778557455\\
22	37.2698109599606\\
23	37.2883942314041\\
24	37.3062105952057\\
25	37.3233299822572\\
26	37.3398121144697\\
27	37.3557085698744\\
28	37.3710643332266\\
29	37.385919182084\\
30	37.4003077940128\\
31	37.4142615838597\\
32	37.4278087953238\\
33	37.4409748914695\\
34	37.4537829739306\\
35	37.4662540805463\\
36	37.4784073588602\\
37	37.4902659205268\\
38	37.5018421652906\\
39	37.5131491048464\\
40	37.5242009969527\\
41	37.5350108559033\\
42	37.5455909641701\\
43	37.5559526992067\\
44	37.5661066158909\\
45	37.5760625632481\\
46	37.5858296238287\\
47	37.5954163241617\\
48	37.6048308558019\\
49	37.6140808144358\\
50	37.6231733549056\\
51	37.6321152166826\\
52	37.6409127042315\\
53	37.6495717925434\\
54	37.6580981368512\\
55	37.666497113163\\
56	37.6747738008016\\
57	37.6829330059971\\
58	37.6909792663729\\
59	37.6989167068383\\
60	37.7067495102959\\
61	37.7144815611477\\
62	37.7221165516014\\
63	37.7296579638888\\
64	37.737109103619\\
65	37.7444731133011\\
66	37.7517529816492\\
67	37.758951552075\\
68	37.7660715307372\\
69	37.7731154939524\\
70	37.7800858963887\\
71	37.7869850768074\\
72	37.7938152623862\\
73	37.8005785762208\\
74	37.807277042655\\
75	37.8139125924951\\
76	37.8204870556818\\
77	37.8270021924798\\
78	37.8334596848536\\
79	37.839861137799\\
80	37.8462080840854\\
81	37.8525019869494\\
82	37.8587472198884\\
83	37.8649430299628\\
84	37.8710895645031\\
85	37.8771881575913\\
86	37.8832400664869\\
87	37.8892464726878\\
88	37.8952084935215\\
89	37.9011271899644\\
90	37.9070035729179\\
91	37.9128386082167\\
92	37.9186332198564\\
93	37.9243882926941\\
94	37.9301046750421\\
95	37.9357831806065\\
96	37.9414245903708\\
97	37.9470296541413\\
98	37.9525990921459\\
99	37.958133596464\\
100	37.9636338323028\\
101	37.9691004393137\\
102	37.9745340328263\\
103	37.9799352048622\\
104	37.9853045253634\\
105	37.9906425432164\\
106	37.9959497869629\\
107	38.0012267662051\\
108	38.0064739722023\\
109	38.0116918787167\\
110	38.016880943003\\
111	38.0220416063946\\
112	38.0271742950815\\
113	38.0322794168341\\
114	38.0373573687437\\
115	38.0424085367653\\
116	38.0474332941301\\
117	38.0524320015083\\
118	38.0574050078278\\
119	38.0623526505496\\
120	38.067275256468\\
121	38.0721731426294\\
122	38.0770466158202\\
123	38.0818959726539\\
124	38.0867215004788\\
125	38.0915234778187\\
126	38.0963021747198\\
127	38.1010578531594\\
128	38.1057907671956\\
129	38.1105011635417\\
130	38.1151892807699\\
131	38.1198552918808\\
132	38.1244994423706\\
133	38.1291220011677\\
134	38.1337231810088\\
135	38.1383031887978\\
136	38.1428622259105\\
137	38.1474004886166\\
138	38.1519181682058\\
139	38.1564154513146\\
140	38.1608925199805\\
141	38.1653495519055\\
142	38.1697867204273\\
143	38.1742041950711\\
144	38.1786021416004\\
145	38.1829807219405\\
146	38.1873400941304\\
147	38.1916804127369\\
148	38.1960018290046\\
149	38.2003044909406\\
150	38.2045885435343\\
151	38.2088541288475\\
152	38.2131013860608\\
153	38.2173304515817\\
154	38.2215414591162\\
155	38.2257345401058\\
156	38.229909823263\\
157	38.2340674350684\\
158	38.2382074998132\\
159	38.2423301394919\\
160	38.2464354740323\\
161	38.2505236214485\\
162	38.2545946975157\\
163	38.2586488165031\\
164	38.2626860907378\\
165	38.2667066309508\\
166	38.2707105456606\\
167	38.2746979422062\\
168	38.2786689261053\\
169	38.2826236017781\\
170	38.2865620719711\\
171	38.2904844377238\\
172	38.2943907990288\\
173	38.2982812547482\\
174	38.3021559018889\\
175	38.3060148365132\\
176	38.3098581538947\\
177	38.3136859478896\\
178	38.3174983109343\\
179	38.321295334404\\
180	38.3250771091478\\
181	38.328843724706\\
182	38.3325952676396\\
183	38.3363318240337\\
184	38.3400534831927\\
185	38.343760331196\\
186	38.3474524533715\\
187	38.3511299340352\\
188	38.3547928571533\\
189	38.3584413050822\\
190	38.3620753601957\\
191	38.3656951032817\\
192	38.3693006140692\\
193	38.3728919708902\\
194	38.3764692482107\\
195	38.3800325242709\\
196	38.3835818752061\\
197	38.3871173736785\\
198	38.3906390958188\\
199	38.3941471176344\\
200	38.3976415144732\\
};
\addlegendentry{CVX \cite{mu2021simultaneously}}

\addplot [color=red, solid, line width=2.0pt]
  table[row sep=crcr]{%
1	6.537783921896\\
2	22.0402108541055\\
3	33.1461682846689\\
4	35.3530168750234\\
5	36.1210751362181\\
6	36.4636589550052\\
7	36.6509156742817\\
8	36.7682793822996\\
9	36.850685265481\\
10	36.9136816891316\\
11	36.9645077809301\\
12	37.0072383788717\\
13	37.0446245603275\\
14	37.0779710946664\\
15	37.1083156680913\\
16	37.1358459724169\\
17	37.1617022809211\\
18	37.1857555806962\\
19	37.2082261382681\\
20	37.2297347447352\\
21	37.2499857508503\\
22	37.2694836424112\\
23	37.2880623625296\\
24	37.3057416818463\\
25	37.3228284441394\\
26	37.3392467978131\\
27	37.355034764156\\
28	37.3702724684445\\
29	37.3849702795765\\
30	37.399283152477\\
31	37.4131737510881\\
32	37.4266204686383\\
33	37.439789724301\\
34	37.4525323393287\\
35	37.4649627375582\\
36	37.4770133108532\\
37	37.4888367038784\\
38	37.5003653407209\\
39	37.5114073588028\\
40	37.5226941332081\\
41	37.5334620326175\\
42	37.5440468267072\\
43	37.5542847152971\\
44	37.5643607872001\\
45	37.5742575217575\\
46	37.5838946231655\\
47	37.5933597978578\\
48	37.602686410083\\
49	37.6117955068771\\
50	37.6207500736356\\
51	37.6295615347768\\
52	37.6382172488337\\
53	37.6467211221884\\
54	37.6550342181968\\
55	37.6632797097737\\
56	37.6713417971818\\
57	37.679249743711\\
58	37.6870058483247\\
59	37.6947403971602\\
60	37.7023125908207\\
61	37.709804969219\\
62	37.7171506148782\\
63	37.7244868332846\\
64	37.7316764017889\\
65	37.7387799944929\\
66	37.7458069378121\\
67	37.7527601954016\\
68	37.7596194617441\\
69	37.7663919814359\\
70	37.7730211989861\\
71	37.779658551545\\
72	37.7862252557702\\
73	37.7927130413504\\
74	37.7991287054051\\
75	37.8054320537837\\
76	37.8116979707545\\
77	37.817951543585\\
78	37.8241296380937\\
79	37.8302369329727\\
80	37.8362356081007\\
81	37.8422570918288\\
82	37.8482384289177\\
83	37.8541477505822\\
84	37.8599103566322\\
85	37.8657287312989\\
86	37.8715273037124\\
87	37.8772262083484\\
88	37.8829376523397\\
89	37.8885650117723\\
90	37.8941644433178\\
91	37.8996203777711\\
92	37.9051218585607\\
93	37.9106659841773\\
94	37.9161388173718\\
95	37.921567895032\\
96	37.9269314256156\\
97	37.9322637970223\\
98	37.9375837653175\\
99	37.9427250441456\\
100	37.9479588271524\\
101	37.953169245905\\
102	37.9583576436114\\
103	37.9635013707529\\
104	37.9686447071854\\
105	37.9737122487078\\
106	37.9787566032213\\
107	37.9837843397223\\
108	37.9888109880413\\
109	37.9937955913107\\
110	37.9987209328792\\
111	38.003640973983\\
112	38.0085374501621\\
113	38.013410392759\\
114	38.0182736072158\\
115	38.022857852332\\
116	38.0276259772995\\
117	38.0324294338838\\
118	38.0372105369702\\
119	38.04194486955\\
120	38.0466666732193\\
121	38.0513478117201\\
122	38.0559773698303\\
123	38.0606463540174\\
124	38.0652558120403\\
125	38.0698456805616\\
126	38.0744106360402\\
127	38.0789764437301\\
128	38.0834979016655\\
129	38.0880046105185\\
130	38.0925097099671\\
131	38.0969501477116\\
132	38.101405953439\\
133	38.1058094729598\\
134	38.1101880707268\\
135	38.1145799973349\\
136	38.1189570578952\\
137	38.1232861110775\\
138	38.1276370873056\\
139	38.1318900556825\\
140	38.1361500122337\\
141	38.1404230937618\\
142	38.1446892295643\\
143	38.1488801468141\\
144	38.1530271465178\\
145	38.1572613951847\\
146	38.161415335467\\
147	38.1655869263894\\
148	38.169660272223\\
149	38.1737962484936\\
150	38.1778626504126\\
151	38.1819196829172\\
152	38.1859940413039\\
153	38.1899786587727\\
154	38.1939312476605\\
155	38.1979780929777\\
156	38.2019369230512\\
157	38.2058550107813\\
158	38.2097467118547\\
159	38.2137345411524\\
160	38.2177037245418\\
161	38.2216431098352\\
162	38.2254251563103\\
163	38.2293810345825\\
164	38.2333165946404\\
165	38.2371887878709\\
166	38.2410330592116\\
167	38.2448530508451\\
168	38.2487030654715\\
169	38.252513841394\\
170	38.2563117228637\\
171	38.2600821410048\\
172	38.2638293067474\\
173	38.2675561769194\\
174	38.2713096837043\\
175	38.2749244267546\\
176	38.2786076379515\\
177	38.2823046901634\\
178	38.2859846990136\\
179	38.2896422949398\\
180	38.2933159767353\\
181	38.2969401948554\\
182	38.3005596862171\\
183	38.3041585008057\\
184	38.3077622529554\\
185	38.3113307383971\\
186	38.3149320843718\\
187	38.3183384224223\\
188	38.3218670302438\\
189	38.3253645905172\\
190	38.3288739223588\\
191	38.332330801335\\
192	38.3357831741468\\
193	38.3392785208174\\
194	38.3427338478836\\
195	38.3461522895305\\
196	38.3495738560613\\
197	38.3529593169748\\
198	38.3562484982255\\
199	38.3596481277776\\
200	38.3630245614433\\
};
\addlegendentry{Element-STAR}

\end{axis}
\end{tikzpicture}%
} %{\input{Convergence}} %{\input{conv_cdf}} %
		\caption{Convergence Plot for $M= 8, K_{\rt} = 3, K_{\ttxt} = 2, N = 40, P_t = 20$~dB }
		\label{convergence}
\end{figure}

\begin{figure}		
		\scalebox{0.80}{\definecolor{mycolor1}{rgb}{0.00000,0.44700,0.74100}%
\definecolor{mycolor2}{rgb}{0.85000,0.32500,0.09800}%
\begin{tikzpicture}

\begin{axis}[%
width=0.9\figW,
height=\figH,
at={(0\figW,0\figH)},
scale only axis,
xmin=0,
xmax=800,
xlabel style={font=\color{white!15!black}},
xlabel={$\bf{Run-Time \:[sec]}$},
ymin=0,
ymax=1,
ylabel style={font=\color{white!15!black}},
ylabel={$\bf{CDF}$},
axis background/.style={fill=white},
xmajorgrids,
ymajorgrids,
legend style={at={(0.99,0.01)}, anchor=south east, legend cell align=left, align=left, draw=white!15!black, row sep=-0.05cm, font = \scriptsize}
]
\addplot [color=red, solid, line width=2.0pt]
  table[row sep=crcr] {%
13.3700147	0\\
13.3700147	0.0102040816326531\\
13.4865572	0.0204081632653061\\
13.4930795	0.0306122448979592\\
13.5305985	0.0408163265306123\\
13.5516496	0.0510204081632654\\
13.5593729	0.0612244897959184\\
13.5767631	0.0714285714285715\\
13.6212608	0.0816326530612246\\
13.6482319	0.0918367346938777\\
13.6841776	0.102040816326531\\
13.7248904	0.112244897959184\\
13.7943624	0.122448979591837\\
13.8548135	0.13265306122449\\
14.0534955	0.142857142857143\\
14.5265506	0.153061224489796\\
16.1790476	0.163265306122449\\
19.9754215	0.173469387755102\\
23.5240373	0.183673469387755\\
24.184846	0.193877551020408\\
24.8998325	0.204081632653061\\
25.0364043	0.214285714285715\\
25.2306568	0.224489795918368\\
25.3446845	0.234693877551021\\
25.5792648	0.244897959183674\\
26.9467656	0.255102040816327\\
27.9985726	0.26530612244898\\
28.1256375	0.275510204081633\\
28.8044066	0.285714285714286\\
29.3011788	0.295918367346939\\
29.6571837	0.306122448979592\\
29.7068522	0.316326530612245\\
29.7498987	0.326530612244898\\
29.7958529	0.336734693877551\\
29.9442033	0.346938775510204\\
30.0796774	0.357142857142858\\
30.2938099	0.367346938775511\\
30.3444433	0.377551020408164\\
30.4155852	0.387755102040817\\
30.4570246	0.39795918367347\\
30.4752967	0.408163265306123\\
30.5414409	0.418367346938776\\
30.562279	0.428571428571429\\
30.7182966	0.438775510204082\\
30.7556553	0.448979591836735\\
30.7871946	0.459183673469388\\
30.8891511	0.469387755102041\\
30.9225364	0.479591836734694\\
30.9846147	0.489795918367347\\
30.999281	0.500000000000001\\
31.0033603	0.510204081632654\\
31.134847	0.520408163265307\\
31.7736791	0.53061224489796\\
33.458054	0.540816326530613\\
33.9242603	0.551020408163266\\
34.0225541	0.561224489795919\\
34.1543416	0.571428571428572\\
36.5161217	0.581632653061225\\
38.1923656	0.591836734693878\\
39.0366601	0.602040816326531\\
39.6054293	0.612244897959184\\
42.8081266	0.622448979591837\\
42.9354907	0.63265306122449\\
43.026475	0.642857142857143\\
43.3419225	0.653061224489796\\
43.5179287	0.663265306122449\\
43.5346959	0.673469387755103\\
43.6664473	0.683673469387756\\
43.7508594	0.693877551020409\\
43.7726376	0.704081632653062\\
43.85887	0.714285714285715\\
43.9090823	0.724489795918368\\
43.9958039	0.734693877551021\\
44.4357429	0.744897959183674\\
44.4361942	0.755102040816327\\
44.5676963	0.76530612244898\\
44.7568076	0.775510204081633\\
44.916185	0.785714285714286\\
44.9848636	0.795918367346939\\
45.0171664	0.806122448979592\\
45.0183965	0.816326530612245\\
45.1103634	0.826530612244898\\
45.4061105	0.836734693877551\\
45.6738558	0.846938775510204\\
45.7036849	0.857142857142857\\
46.1117988	0.86734693877551\\
46.1778435	0.877551020408163\\
46.3048795	0.887755102040816\\
46.4201971	0.897959183673469\\
47.1173486	0.908163265306122\\
47.6502213	0.918367346938776\\
49.7194573	0.928571428571429\\
49.8622035	0.938775510204082\\
51.4398378	0.948979591836735\\
51.6192829	0.959183673469388\\
58.3093757	0.969387755102041\\
64.1059172	0.979591836734694\\
64.7451011	0.989795918367347\\
67.966187	1\\
};
\addlegendentry{Element-STAR}

\addplot [color=mycolor1, solid, line width=2.0pt]
  table[row sep=crcr]{%
203.6903028	0\\
203.6903028	0.0102040816326531\\
203.7013965	0.0204081632653061\\
203.7707242	0.0306122448979592\\
203.811114	0.0408163265306123\\
203.8320146	0.0510204081632654\\
203.9198624	0.0612244897959184\\
204.2683002	0.0714285714285715\\
204.654314	0.0816326530612246\\
204.9454905	0.0918367346938777\\
204.9501575	0.102040816326531\\
205.1566563	0.112244897959184\\
205.283472	0.122448979591837\\
206.8063686	0.13265306122449\\
210.6028236	0.142857142857143\\
%%175.6028236	0.14857142857143\\
%%180.6028236	0.1498957142857143\\
%%185.6028236	0.15098957142857143\\
%%190.6028236	0.15198957142857143\\
%180.6028236	0.142857142857143\\
%%197.1968715	0.153061224489796\\
%%204.0063775	0.163265306122449\\
%%210.0063775	0.169265306122449\\ %
229.1340186	0.173469387755102\\
229.8577272	0.183673469387755\\
231.8009545	0.193877551020408\\
231.9562717	0.204081632653061\\
232.484313	0.214285714285715\\
232.9391895	0.224489795918368\\
246.4988386	0.234693877551021\\
248.3075811	0.244897959183674\\
249.1329105	0.255102040816327\\
251.9655253	0.26530612244898\\
252.2882009	0.275510204081633\\
253.299432	0.285714285714286\\
254.2049671	0.295918367346939\\
256.6593398	0.306122448979592\\
258.0810875	0.316326530612245\\
262.3945681	0.326530612244898\\
263.7401515	0.336734693877551\\
269.1951068	0.346938775510204\\
269.60576	0.357142857142858\\
269.7407185	0.367346938775511\\
269.8454131	0.377551020408164\\
269.9991959	0.387755102040817\\
270.0521553	0.39795918367347\\
270.1726683	0.408163265306123\\
270.414676	0.418367346938776\\
270.6622569	0.428571428571429\\
270.6835262	0.438775510204082\\
270.8086187	0.448979591836735\\
271.1405827	0.459183673469388\\
271.7149188	0.469387755102041\\
273.466088	0.479591836734694\\
277.057343	0.489795918367347\\
278.2767032	0.500000000000001\\
279.7908787	0.510204081632654\\
281.0377274	0.520408163265307\\
284.6054315	0.53061224489796\\
288.7103128	0.540816326530613\\
289.0213936	0.551020408163266\\
291.9955507	0.561224489795919\\
300.2260052	0.571428571428572\\
301.4270847	0.581632653061225\\
309.4270847	0.5891632653061225\\ %
311.6077898	0.591836734693878\\
315.6077898	0.6001836734693878\\%
321.6077898	0.60191836734693878\\%
329.6077898	0.60193836734693878\\%
%%334.2977962	0.602040816326531\\
%%334.4187608	0.612244897959184\\
%%334.6147155	0.622448979591837\\
%%334.7348548	0.63265306122449\\
%%334.7598235	0.642857142857143\\
%%334.8116826	0.653061224489796\\
%%335.2364165	0.663265306122449\\
%%335.252905	0.673469387755103\\
%%335.2831668	0.683673469387756\\
%%335.5699552	0.693877551020409\\
%%335.7359887	0.704081632653062\\
%%336.1235729	0.714285714285715\\
%%336.4124469	0.724489795918368\\
%%336.8630186	0.734693877551021\\
%%337.1911489	0.744897959183674\\
%%337.578115	0.755102040816327\\
%%337.7148954	0.76530612244898\\
%%337.9717352	0.775510204081633\\
%%338.0624968	0.785714285714286\\
%%338.109838	0.795918367346939\\
%338.196594	0.806122448979592\\
%338.2808941	0.816326530612245\\
%%348.196594	0.806122448979592\\ %
%%358.2808941	0.816326530612245\\%
%%368.2808941	0.8169326530612245\\%
%%378.2808941	0.818326530612245\\%
%%387.3684494	0.826530612244898\\
%%394.5690723	0.836734693877551\\
%%399.0685398	0.846938775510204\\
406.4414838	0.857142857142857\\
436.9218252	0.86734693877551\\
449.1778485	0.877551020408163\\
459.9190595	0.887755102040816\\
492.2453425	0.897959183673469\\
546.0945198	0.908163265306122\\
624.2000525	0.918367346938776\\
695.5965368	0.928571428571429\\
695.7503107	0.938775510204082\\
696.1913358	0.948979591836735\\
699.6019573	0.959183673469388\\
700.6602834	0.969387755102041\\
708.9434822	0.979591836734694\\
709.1544599	0.989795918367347\\
800.8508107	1\\
};
\addlegendentry{CVX \cite{mu2021simultaneously}}
\end{axis}

\end{tikzpicture}%
} %{\input{Convergence}} %{\input{conv_cdf}} %
		\caption{CDF plot of the runtime needed until convergence for $M= 8, K_{\rt} = 3, K_{\ttxt} = 2, N = 40, P_t = 10$~dB  }
		\label{runtime}
\end{figure}
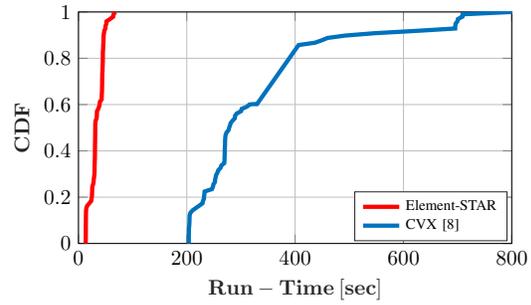

The complexity of the proposed Element-STAR algorithm is dominated by computing $\bfw_k$ in each iteration by \eqref{solw} and updating $a_{\rt, n}$, $a_{\ttxt, n}$ by \eqref{15} and \eqref{16} $N$ times in each iteration, resulting in an overall complexity of $\Ocal \left(I_1  \left(K N M + K N^2 + I_{2} N^2 \right) + I_1 K M^3 \right)$, where $I_1$ is the total number of iterations required for the algorithm to converge and $I_2$ is the number of iterations required for the iterative sub-problem for the update of the passive beamforming of the STAR-RIS. On the other hand, the complexity of the CVX-based algorithm of \cite{mu2021simultaneously} is given by $\Ocal \left(I_0 \left( K N M + K N^2 +  I_3 N^{3.5} \right) +  I_0 K M^3 \right)$, where $I_0$ and $I_3$ are the total number of iterations required for the convergence of the entire algorithm and the interior-point solver respectively. The convergence plot in Fig.~\ref{convergence} for $P_t$ = 20 dB illustrates that the proposed algorithm converges to a similar value as the CVX approach despite being computationally much less expensive. We also compare the runtime of the proposed algorithm with that of the CVX-based algorithm \cite{mu2021simultaneously} in Fig.~\ref{runtime}, where the cumulative distribution function (CDF) of the runtime needed until convergence is plotted for $P_{\text{t}} = 10$ dB. It is evident from the figure that the runtime of the proposed algorithm is at least an order of magnitude smaller than that of \cite{mu2021simultaneously}. 

\section{Results and Discussion}
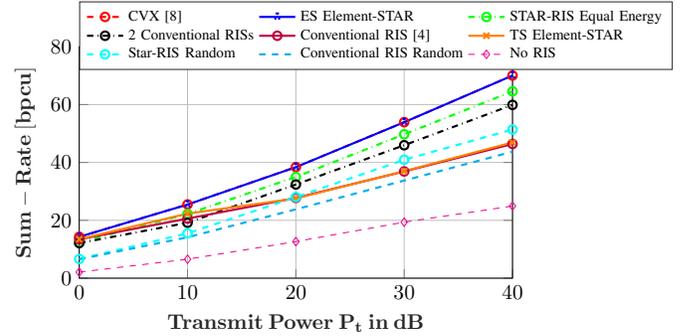
\begin{figure}		
		\scalebox{0.80}{% This file was created by matlab2tikz.
%
%The latest updates can be retrieved from
%  http://www.mathworks.com/matlabcentral/fileexchange/22022-matlab2tikz-matlab2tikz
%where you can also make suggestions and rate matlab2tikz.
%
\definecolor{mycolor1}{rgb}{1.00000,1.00000,0.00000}%
\definecolor{mycolor2}{rgb}{1.00000,0.00000,1.00000}%
\definecolor{mycolor3}{rgb}{0.00000,1.00000,1.00000}%
\begin{tikzpicture}

\begin{axis}[%
width=0.9\figW,
height=\figH,
at={(0\figW,0\figH)},
scale only axis,
xmin=0,
xmax=40,
xlabel style={font=\color{white!15!black}},
xlabel={$\bf{Transmit\:Power\:P_t\:in\:dB}$},
ymin=0,
ymax=80,
ylabel style={font=\color{white!15!black}},
ylabel={$\bf{Sum-Rate\:[bpcu]}$},
axis background/.style={fill=white},
xmajorgrids,
ymajorgrids,
legend style={at={(0,0.90)}, anchor=south west, legend cell align=left, align=left, draw=white!15!black, row sep=-0.05cm, font=\scriptsize},
legend columns=3
]

\addplot [color=red, dashed, line width=1.0pt, mark=o, mark options={solid, red}]
  table[row sep=crcr]{%
0	14.2872452831077\\
10	25.4718996369947\\
20	38.3976415144733\\
30	53.9105546038364\\
40	70.0370684037569\\
};
\addlegendentry{CVX \cite{mu2021simultaneously}}

\addplot [color=blue, line width=1.0pt, mark=diamond, mark options={dotted, blue}]
  table[row sep=crcr]{%
0	14.2834096066431\\
10	25.4474824034825\\
20	38.3630245614433\\
30	53.9052343604672\\
40	70.0329028541184\\
};
\addlegendentry{ES Element-STAR}

\addplot [color=green, dashdotted, line width=1.0pt, mark=o, mark options={solid, green}]
  table[row sep=crcr]{%
0	12.3772399385275\\
10	22.1905957734855\\
20	34.9810785114522\\
30	49.7242268238192\\
40	64.5706709884458\\
};
\addlegendentry{STAR-RIS Equal Energy}

\addplot [color=black, dashdotted, line width=1.0pt, mark=o, mark options={solid, black}]
  table[row sep=crcr]{%
0	12.1171571853821\\
10	19.1610803062641\\
20	32.3813758223211\\
30	45.9494645937451\\
40	59.8927814015105\\
};
\addlegendentry{2 Conventional RISs}

\addplot [color=purple, solid, line width=1.0pt, mark=o, mark options={solid, purple}]
   table[row sep=crcr]{%
0	13.4392377669557\\
10	20.5957476237145\\
20	27.7380506893549\\
30	36.8910149428552\\
40	46.3843400606471\\
};
\addlegendentry{Conventional RIS \cite{guo2020weighted}}

\addplot [color=orange, solid, line width=1.0pt,  mark=x, mark options={solid, orange}]
   table[row sep=crcr]{%
0	13.481783189179222\\
10	22.318960439489760\\
20	27.548646270975205\\
30	36.988659026432850\\
40	46.921224426095640\\
};
\addlegendentry{TS Element-STAR}

\addplot [color=mycolor3, dashed, line width=1.0pt, mark=o, mark options={solid, mycolor3}]
  table[row sep=crcr]{%
0	6.61854546193057\\
10	15.4909019122051\\
20	28.0210117385178\\
30	40.9209185734135\\
40	51.3853469403499\\
};
\addlegendentry{Star-RIS Random}

\addplot [color=cyan, dashed, line width=1.0pt]
  table[row sep=crcr]{%
0	6.53676075728241\\
10	14.1389201020832\\
20	23.7923588597097\\
30	33.7443063824002\\
40	43.711571975223\\
};
\addlegendentry{Conventional RIS Random}

\addplot [color=magenta, dashed, mark=diamond, mark options={solid,magenta}]
  table[row sep=crcr]{%
0	2.0901853373541\\
10	6.55356438174049\\
20	12.6520852698751\\
30	19.3678893580872\\
40	24.9339273473581\\
};
\addlegendentry{No RIS}

\end{axis}

\end{tikzpicture}%
} %{\input{Convergence}} %{\input{conv_cdf}} %
		\caption{Sum-Rate vs Transmit Power $P_t$ in dB }
		\label{sum-rate}
\end{figure}

\begin{figure}		
		\scalebox{0.80}{% This file was created by matlab2tikz.
%
%The latest updates can be retrieved from
%  http://www.mathworks.com/matlabcentral/fileexchange/22022-matlab2tikz-matlab2tikz
%where you can also make suggestions and rate matlab2tikz.
%
\definecolor{mycolor1}{rgb}{1.00000,1.00000,0.00000}%
\definecolor{mycolor2}{rgb}{1.00000,0.00000,1.00000}%
\definecolor{mycolor3}{rgb}{0.00000,1.00000,1.00000}%
\begin{tikzpicture}

\begin{axis}[%
width=0.9\figW,
height=\figH,
at={(0\figW,0\figH)},
scale only axis,
xmin=0,
xmax=100,
xlabel style={font=\color{white!15!black}},
xlabel={$\bf{\it{N}}$},
ymin=20,
ymax=62,
ylabel style={font=\color{white!15!black}},
ylabel={$\bf{Sum-Rate\:[bpcu]}$},
axis background/.style={fill=white},
xmajorgrids,
ymajorgrids,
legend style={at={(0,1.0)}, anchor=south west, legend cell align=left, align=left, draw=white!15!black, row sep=-0.05cm, font=\scriptsize},
legend columns=3
]

\addplot [color=red, dashed, line width=1.0pt, mark=o, mark options={solid, red}] 
table[row sep=crcr]{%
0   24.2780723234814\\
20	47.78635085720986\\
60	55.7384013705109\\
100	60.1328417230858\\
};
\addlegendentry{CVX \cite{mu2021simultaneously}}

\addplot [color=blue, line width=1.0pt, mark=diamond, mark options={dotted, blue}]
  table[row sep=crcr]{%
 0   24.2780723234814\\
20	47.7635085720986\\
60	55.6384013705109\\
100	60.0328417230858\\
};
\addlegendentry{ES Element-STAR}

\addplot [color=green, dashdotted, line width=1.0pt, mark=o, mark options={solid, green}]
  table[row sep=crcr]{%
0   24.2780723234814\\
20	44.5257490569603\\
60	52.6717997219073\\
100	54.2476109863119\\
};
\addlegendentry{STAR-RIS Equal Energy}

\addplot [color=black, dashdotted, line width=1.0pt, mark=o, mark options={solid, black}]
  table[row sep=crcr]{%
 0   24.2780723234814\\
20	43.4574728478981\\
60	50.7003416745951\\
100	52.3099634330734\\
};
\addlegendentry{2 Conventional RISs}

\addplot [color=purple, solid, line width=1.0pt, mark=o, mark options={solid, purple}]
  table[row sep=crcr]{%
0   24.2780723234814\\
20	33.078975731365\\
60	39.760041181411\\
100	42.3699124836914\\
};
\addlegendentry{Conventional RIS \cite{guo2020weighted}}

\addplot [color=orange, solid, line width=1.0pt,  mark=x, mark options={solid, orange}]
 table[row sep=crcr]{%
 0   24.2780723234814\\
20	32.9812998809325\\
60	39.9443326718025\\
100	42.7516706808751\\
};
\addlegendentry{TS Element-STAR}

\addplot [color=mycolor3, dashed, line width=1.0pt, mark=o, mark options={solid, mycolor3}]
  table[row sep=crcr]{%
  0   24.2780723234814\\
20	36.1244624313904\\
60	45.4798418228585\\
100	46.15013186836655\\
};
\addlegendentry{Star-RIS Random}

\addplot [color=cyan, dashed, line width=1.0pt]
  table[row sep=crcr]{%
  0   24.2780723234814\\
20	32.0809636238175\\
60	35.5333491272259\\
100	36.2801503993922\\
};
\addlegendentry{Conventional RIS Random}

\addplot [color=magenta, mark=diamond, mark options={solid,magenta}]
  table[row sep=crcr]{%
  0   24.2780723234814\\
20	24.2780723234814\\
60	24.4586196798122\\
100	24.5681311703995\\
};
\addlegendentry{No RIS}

\end{axis}

\end{tikzpicture}%
} %{\input{Convergence}} %{\input{conv_cdf}} %
		\caption{Sum-Rate vs $N$ at $P_t = 30$ dB }
		\label{N}
\end{figure}
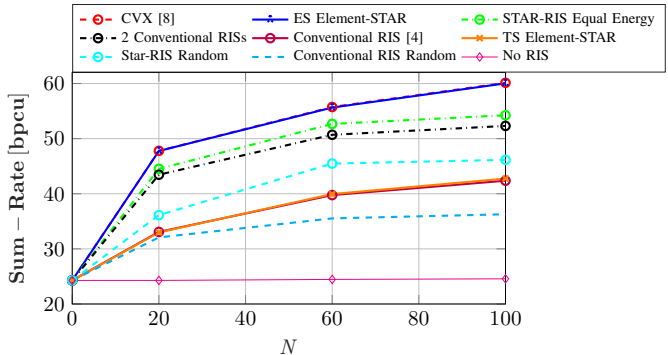

In this section, numerical results are provided to validate the effectiveness of the proposed algorithm. The channels are generated according to the 3GPP specifications \cite{etsi5138}. The details of the channel modelling are given in \cite{syed2024design}. The BS is placed at (0,0,0) $\text{m}$ with $M = 8$ antennas and the STAR-RIS at (50,10,0) $\text{m}$ with $N = 40$ elements. We consider $K = 5$ single-antenna users with $K_{\rt} = 3$ reflecting users placed at $(50,0,0) \text{m}$, $(55,0,0) \text{m}$ and $(53,5,0) \text{m}$, and $K_{\ttxt} = 2$ transmitting users placed at $(50,20,0) \text{m}$ and $(55,15,0) \text{m}$. The sum-rate of the users, which is averaged over 100 different realisations of the channels is taken as the performance metric. The noise variance $\sigma_k^2$ is set to 1 for all users and the average sum-rate is plotted w.r.t. the transmit power levels $P_t$ in Fig.~\ref{sum-rate}.   
We compare both the ES and TS modes of STAR-RIS and compare them with the conventional RIS. The figure illustrates that the proposed low-complexity Element-Star algorithm performs similar to the CVX-based algorithm of \cite{mu2021simultaneously}. Moreover, we achieve a considerable performance gain by optimising the amplitude coefficients of reflection and transmission, instead of using uniform energy splitting (given by {\it{STAR-RIS Equal Energy}}), where half of the energy of the incident signal is used for reflection and the other half is used for transmission mode for each element of the STAR-RIS. The curve {\it{2 Conventional RISs}} considers the case in which 2 reflecting-only RISs are stacked together, one serving the users on the reflecting side and the other serving those on the transmitting side. For a fair comparison, each of the 2 RISs is equipped with $N/2$ elements. This is a special case of MS with half of the elements used for reflecting mode and the other half for transmitting mode. The MS mode cannot achieve the similar gain as the ES, as discussed before and it is also evident from the plot that there is a significant reduction in the achievable sum-rate compared to the ES mode. The {\it{STAR-RIS Random}} approach employs STAR-RIS with equal amplitude coefficients and random phase shifts for all the elements. Additionally, it can be seen from Fig.~\ref{sum-rate} that the TS mode of the STAR-RIS is unable to improve the sum-rate of the users compared to the conventional RIS of \cite{guo2020weighted} (except at very low $P_t$), where all the $N$ elements of the reflecting-only RIS only serve the 3 reflecting users. This is attributed to the fact that the reflecting-only RIS serves 3 reflecting users at all time, whereas the TS STAR-RIS serves 3 users for some time and then 2 transmitting users in the remaining time of each time slot. Hence, it is unable to bring any advantage compared to the conventional RIS at high $P_t$ in terms of the sum-rate, but the TS mode ensures that all of the 5 users are served for some time in each time slot. Finally, we plot the sum-rate vs $N$ in Fig.~\ref{N}.
The plot shows that the optimisation of the amplitude coefficients of the STAR-RIS leads to an increase in the gain w.r.t. $N$ compared to the {\it{Equal Energy STAR-RIS}}. As expected, the slopes of the random RISs are lower than their optimised counterparts. We also observe that the slope of the TS mode is similar to that of the conventional RIS and it does not improve the sum-rate compared to the conventional RIS. Moreover, the slope of {\it{2 Conventional RISs}} is lower than that of the ES mode of STAR-RIS as each RIS there consists of $N/2$ elements and is used for either reflection or transmission. %On the other hand, by properly optimising the energy split between reflection and transmission modes for each element of the STAR-RIS as in the ES protocol,   
\section{Conclusion}
In this work, we have proposed a low complexity solution for the design of the STAR-RIS. We have analysed the ES and the TS protocols, and the simulation results illustrate that the ES protocol brings a significant performance gain compared to the conventional RIS. However, the TS protocol does not improve the achievable sum-rate of the users compared to the reflecting-only RIS at high tranmsit power levels. Nonetheless, it ensures that all users on both sides of the STAR-RIS are at least served for some time.
\appendix
\begin{proof}
\eqref{P5} can be reformulated by the FP approach as a function of $\delta$ and ignoring the constant terms as $t(\delta) = a \:\delta - b \:\delta^2$, where $a = c_{\rt} - c_{\ttxt} + 2 d_{\ttxt} $ and $b = d_{\rt} + d_{\ttxt}$ with 
{\small
\begin{align*}
    c_{\rt /  \ttxt} =  &\sum\nolimits_{k=1}^{K_{\rt /\ttxt }}\Big(\log(1 + \lambda^{\rt /  \ttxt}_k) - \lambda^{\rt /  \ttxt}_k + 2  \sqrt{1 + \lambda^{\rt /  \ttxt}_k} \nonumber \\ \times &\text{Re}\big\{\beta_k^{*\rt /  \ttxt}(\bfh^{\rt /  \ttxt}_k)^{\Hm}\bfw^{\rt /  \ttxt}_k\big\} \Big)  -\sum\nolimits_{k=1}^{K_{\rt /\ttxt }}|\beta^{\rt /  \ttxt}_k|^2 {\sum\nolimits_{{i = 1}}^{K_{\rt /\ttxt }}\big|{(\bfh^{\rt /  \ttxt}_{k})^{\Hm}}{\bfw^{\rt /  \ttxt}_{i}}\big|^2 } \\
     d_{\rt /  \ttxt} &= \sum\nolimits_{k=1}^{K_{\rt /\ttxt }} |\beta^{\rt /  \ttxt}_k|^2 \sigma_k^2.
\end{align*}
}
It is clear that $b \geq 0$, and hence, $t(\delta)$ is concave in $\delta$. 
\end{proof}
%\begin{align*}
%    d_{\rt /  \ttxt} = \sum\nolimits_{k=1}^{K_{\rt /\ttxt }} |\beta_k|^2 \sigma_k^2.
%\end{align*}

\bibliographystyle{IEEEtran}
\bibliography{bibliography}

\end{document}